\documentclass[pra,twocolumn,showkeys,showpacs,superscriptaddress,floatfix]{revtex4-2}

\usepackage[T1]{fontenc}
\usepackage{epsfig,psfrag}
\usepackage{comment}
\usepackage{latexsym}
\usepackage{graphicx}
\usepackage{dcolumn}
\usepackage{amssymb}
\usepackage{amsmath}
\usepackage{amsfonts}
\usepackage{bm}
\usepackage{mathrsfs} 
\usepackage{bigints}
\usepackage{upgreek}
\usepackage{color}
\usepackage{longtable}
\usepackage{xspace}
\usepackage[squaren]{SIunits} 
\usepackage{epstopdf}
\usepackage{float}
\usepackage[normalem]{ulem}
\usepackage{hyperref}
\usepackage{bbold}
\usepackage{bbm}
\usepackage{tikz}
\usepackage{url}
\usepackage{parskip}
\usepackage{subfigure}
\usepackage[utf8]{inputenc}

\usepackage{filemod}

\newcommand{\eq}[1]{Eq.~\eqref{eq:#1}}

\newcommand{\fig}[1]{Fig.~\ref{fig:#1}}

\newcommand{\braket}[1]{\left \langle {#1} \right \rangle}
\newcommand{\ket}[1]{\left| {#1} \right \rangle}

\begin{document}

\title{Experimental realization of a SU(3) color-orbit coupling in an ultracold gas} 

\author{Chetan S. Madasu}
\altaffiliation[Present address: ]{Atomionics Pte. Ltd., Singapore 618494}
\affiliation{Nanyang Quantum Hub, School of Physical and Mathematical Sciences, Nanyang Technological University, 21 Nanyang Link, Singapore 637371, Singapore}
\affiliation{MajuLab, International Joint Research Unit IRL 3654, CNRS, Universit\'e C\^ote d'Azur, Sorbonne Universit\'e, National University of Singapore, Nanyang
Technological University, Singapore}
\affiliation{Centre for Quantum Technologies, National University of Singapore, 117543 Singapore, Singapore}
\author{Chirantan Mitra}
\affiliation{Nanyang Quantum Hub, School of Physical and Mathematical Sciences, Nanyang Technological University, 21 Nanyang Link, Singapore 637371, Singapore}
\affiliation{MajuLab, International Joint Research Unit IRL 3654, CNRS, Universit\'e C\^ote d'Azur, Sorbonne Universit\'e, National University of Singapore, Nanyang
Technological University, Singapore}
\author{Lucas Gabardos}
\affiliation{Nanyang Quantum Hub, School of Physical and Mathematical Sciences, Nanyang Technological University, 21 Nanyang Link, Singapore 637371, Singapore}
\affiliation{MajuLab, International Joint Research Unit IRL 3654, CNRS, Universit\'e C\^ote d'Azur, Sorbonne Universit\'e, National University of Singapore, Nanyang
Technological University, Singapore}
\author{Ketan D. Rathod}
\altaffiliation[Present address: ]{Bennett University, Greater Noida 201310, India}
\affiliation{Centre for Quantum Technologies, National University of Singapore, 117543 Singapore, Singapore}
\affiliation{MajuLab, International Joint Research Unit IRL 3654, CNRS, Universit\'e C\^ote d'Azur, Sorbonne Universit\'e, National University of Singapore, Nanyang
Technological University, Singapore}
\author{Thomas Zanon-Willette}
\affiliation{Sorbonne Universit\'e CNRS, MONARIS, UMR 8233, F-75005 Paris, France}
\author{Christian Miniatura}
\affiliation{Universit\'e C\^ote d'Azur, CNRS, INPHYNI, 17 rue Julien Laupr\^etre, 06200 Nice, France}
\affiliation{Centre for Quantum Technologies, National University of Singapore, 117543 Singapore, Singapore}
\author{Fr\'ed\'eric Chevy}
\affiliation{Laboratoire de Physique de l'\'Ecole normale sup\'erieure, ENS, Universit\'e PSL, CNRS, Sorbonne Universit\'e, Universit\'e de Paris, F-75005 Paris, France}
\affiliation{Institut Universitaire de France (IUF), 75005 Paris, France}
\author{Chang Chi Kwong}
\affiliation{Nanyang Quantum Hub, School of Physical and Mathematical Sciences, Nanyang Technological University, 21 Nanyang Link, Singapore 637371, Singapore}
\affiliation{MajuLab, International Joint Research Unit IRL 3654, CNRS, Universit\'e C\^ote d'Azur, Sorbonne Universit\'e, National University of Singapore, Nanyang
Technological University, Singapore}
\author{David Wilkowski}
\email[Correspondence: ]{david.wilkowski@ntu.edu.sg}
\affiliation{Nanyang Quantum Hub, School of Physical and Mathematical Sciences, Nanyang Technological University, 21 Nanyang Link, Singapore 637371, Singapore}
\affiliation{MajuLab, International Joint Research Unit IRL 3654, CNRS, Universit\'e C\^ote d'Azur, Sorbonne Universit\'e, National University of Singapore, Nanyang
Technological University, Singapore}
\affiliation{Centre for Quantum Technologies, National University of Singapore, 117543 Singapore, Singapore}


\begin{abstract} 
Spin-orbit interaction couples the spin of a particle to its motion and leads to spin-induced transport phenomena such as spin-Hall effects and Chern insulators.
In this work, we extend the concept of internal-external state coupling to higher internal symmetry, exploring features beyond the established spin-orbit regime. We couple suitable resonant laser beams to a gas of ultracold atoms, thereby inducing artificial SU(3) non-Abelian gauge fields that act on a degenerate ground state manifold comprised of three dark states. We demonstrate the inherent all-state connectivity of SU(3) systems by performing targeted geometric transformations. Then, we investigate color-orbit coupling, an extension of SU(2) spin-orbit coupling to SU(3) systems. We reveal a rich dynamical interplay between three distinct oscillation frequencies, which possesses interesting analogies with neutrino oscillations and quark mixing mechanisms. In the future, the system should provide a testbed for further investigation of topological properties of SU(3) systems.
\end{abstract}

\maketitle 

\noindent {\Large\textbf{Introduction}}

Symmetries play a central role in modern physics \cite{Gross1996}. Their existence enforces mathematical structures, groups of transformations, and conservation laws that govern the evolution of physical systems. For instance, relativistic invariance gives rise to Minkowski spacetime, a causal structure, and the Poincar\'e group \cite{ohlsson2011}. This symmetry also guided Dirac in formulating his equation for the electron \cite{dirac1927quantum}, leading to the discovery of antimatter. 
The concept of gauge invariance, first articulated in classical electrodynamics and further developed in quantum theory, underscores the significance of symmetries. 
Moreover, all known fundamental interactions arise from gauge theories, with Lie groups such as U(1), SU(2), and SU(3) describing the electromagnetic, weak, and strong forces, respectively \cite{braibant2011}.

In contrast to U(1), SU(N) symmetries, with $N>1$, are non-Abelian and accommodate surprising topological properties, for example gapless quantum spin Hall edge states in topological insulators \cite{konig2007quantum}, and the yet-to-be observed magnetic monopoles in particle physics \cite{t1974magnetic}. Over the last decade, quantum simulators have been proposed to explore these exotic systems \cite{banerjee2013, zohar2013, tagliacozzo2013simulation, zhang2014}. In particular, ultracold atomic gases dressed with laser light have allowed to explore SU(N) artificial gauge fields and topological invariants associated to their Berry connection \cite{dalibard2011colloquium, goldman2014light,zhai2015degenerate,aidelsburger2015measuring,zhang2019recent,cooper2019topological}. For SU(2) symmetric systems, the gauge field structure can generally be reduced to a spin-orbit coupling term at the origin of spin Hall  \cite{beeler2013spin,hasan2022wavepacket} and \textit{Zitterbewegung}-like effects in one dimension \cite{PhysRevLett.105.143902,gerritsma2010quantum,PhysRevA.88.021604,leblanc2013direct} or two dimensions \cite{hasan2022wavepacket}. Further theoretical efforts have concentrated on gauge fields associated with larger symmetries, such as SU(3), addressing phenomena that extend beyond the realm of condensed-matter physics. These investigations include color-driven topological phases \cite{PhysRevA.98.043614, PhysRevB.101.245159, condmat1010002, PhysRevA.103.043302, zhou20223}, color superfluidity \cite{PhysRevA.97.023632,Wang2020}, chiral spin texture \cite{grass2014spiral}, and color Hall effect \cite{PhysRevA.106.053313}. Here, the \textit{color} terminology is borrowed from the SU(3) color charge of quarks in high-energy physics \cite{greiner2007}. 
 On the experimental side, the manipulation of quantum systems with underlying SU(3) symmetry has been limited to quantum information \cite{PhysRevLett.120.180401,PhysRevLett.126.210504,PhysRevLett.129.100501,PhysRevApplied.19.034089} and sensing \cite{barfussphasecontrolled2018} applications, without coupling to the system external degree of freedom.
\begin{figure*}[htpb]
\centering
\includegraphics[width=1\textwidth]{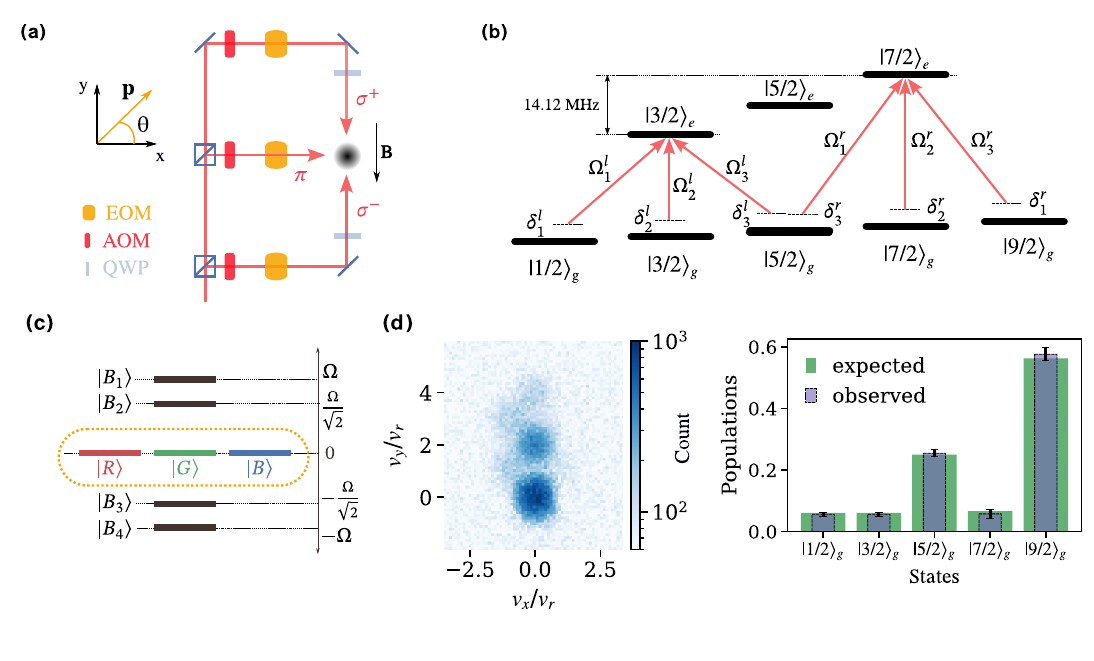}
\caption{\textbf{Experimental setup}. \textbf{a.} A laser beam (red lines), lying in the ($x,y$)-plane, is split into three parts and redirected on a cold atomic cloud of strontium atoms (grey disk). The polarization of each beam is indicated on the graph. Electro- and acousto-optic modulators (EOM, AOM) and quarter-wave plates (QWP) control each beam frequency, Rabi frequency, and polarization state.  A magnetic field bias $\textit{B}=67\,$G is applied along the $y$-axis. The in-plane atomic mean momentum \textbf{p} and its polar angle $\theta$ are defined in the laser beams reference frame (see Methods). 
\textbf{b.} 
The red arrows show the quasi-resonant atomic transitions of the intercombination line driven by the laser beams. These transitions form a 2-tripod scheme made of a left and a right tripod configuration sharing the common ground state $\ket{5/2}_g$. The 2-tripod laser Rabi frequencies and single photon detunings are $\Omega^{l, r}_{1,2,3}$ and $\delta^{l, r}_{1,2,3}$, respectively. Here, all the Rabi frequencies are equal, namely $\Omega^{l,r}_{1,2,3} = \Omega$.
\textbf{c.} The bright ($\ket{\textrm{B}}_j$, $j=1,2,3,4$) and color ($\ket{R}$, $\ket{B}$, $\ket{G}$) dressed states diagonalize the atom-laser coupling Hamiltonian in the rotating wave
approximation. At resonance ($\delta^{l, r}_{1,2,3} = 0$), the color states are degenerated with a zero eigenenergy and separated from the bright states by a frequency shift of the scale $\Omega$. For moving atoms and quasi-resonant beams, the color states are quasi-degenerate as long as the Doppler effect and detunings are much smaller than the Rabi frequency. 
Then, the atomic dynamics constrained to the color space are described by an SU(3) gauge field. 
\textbf{d.} Left panel: Time-of-flight (TOF) image showing the measured velocity distribution (in units of the recoil velocity $v_r$) of the atoms in the different ground states after the initial color state preparation. Right panel: From the TOF image, we extract the populations in the different ground states (blue vertical bars) and compare them to the populations expected from the targeted initial state (green vertical bars). The black error bars are the experimental standard deviation. The fidelity of our state preparation protocol is $\mathcal{F}= 0.97(2)$.}
\label{fig:fig1} 
\end{figure*} 

In this work, we explore the wavepacket dynamics of ultracold atoms subject to a homogeneous SU(3) color-orbit synthetic gauge field. When the gauge field connects each color state to its two other color partners, the time evolution of the system exhibits color oscillations reminiscent of quark-color change in QCD \cite{greiner2007}, and similar to flavor oscillation of neutrinos \cite{gonzalez2008phenomenology}. These temporal oscillations involve three different frequencies, a smoking gun of the SU(3) nature of the gauge field. In contrast, the same three-level system subject to an SU(2) gauge field would evolve with at most two oscillation frequencies (see Methods), and it further reduces to one oscillation frequency for a single tripod configuration associated with only two dark states \cite{hasan2022wavepacket}.

\vspace{0.3cm}
\noindent {\Large\textbf{Results}}

\textbf{Experimental system and color states: } Our experiment starts with a degenerate Fermi gas of $N=5.0(5)\times10^4$ strontium atoms ($^{87}$Sr) at a temperature $T=0.26(2) T_F$, where $T_F=211\,$nK is the Fermi temperature \cite{yang2015high,Hasan2022}. All atoms are initially prepared in the internal ground state $|9/2\rangle_g \equiv |\,^1S_0,F=9/2,m_F=9/2\rangle$. A synthetic SU(3) gauge field is produced by shining the atoms with three polarized coplanar laser beams which are quasi-resonant with the intercombination line $\,^1S_0, F=9/2\rightarrow\,^3P_1, F=9/2$ at $689\,$nm (linewidth $\Gamma/2\pi=7.5\,$kHz), see \fig{fig1}a. A magnetic field of $67\,$G, defining the quantization axis, lifts the degeneracy between excited states by approximately $10^3\Gamma$ (Land\'e factor $g_e=2/33$), while the ground states are weakly sensitive to the magnetic field since the Land\'e factor is $g_g=-1.3\times 10^{-4}$. By appropriately choosing the laser detunings and the polarization states according to the selection rules, one realizes a 2-tripod configuration consisting of a left (l) and a right (r) laser tripod sharing a common ground state. The l-tripod connects the ground states $\ket{1/2}_g$, $\ket{3/2}_g$, and $\ket{5/2}_g$ to the excited state $|3/2\rangle_e \equiv |\,^3P_1,F=9/2,m_F=3/2\rangle$ while the r-tripod connects the ground states $\ket{5/2}_g$, $\ket{7/2}_g$, and $\ket{9/2}_g$ to the excited state $|7/2\rangle_e$, see \fig{fig1}b. Within this 2-tripod scheme, only six quasi-resonant transitions are truly effective and all the other off-resonant transitions are discarded. Correspondingly, the excited and ground states left aside by the lasers remain spectators. 

Very generally, this 2-tripod system admits four non-degenerate AC-Stark-shifted dressed states $\ket{\textrm{B}_j}$ ($j=1,2,3,4$), called bright states, and three degenerate states at zero energy, uncoupled to the laser fields and called dark states. They are obtained by diagonalizing the atom-laser coupling Hamiltonian in the rotating wave approximation \cite{hu2014u}, see \fig{fig1}c. The SU(3) color states $\ket{R}$, $\ket{G}$ and $\ket{B}$ that we use hereafter are particular basis vectors in this degenerate dark-state manifold that we now call the color-manifold (see Methods). 
Because they involve excited states, the bright states are prone to spontaneous emission and thus short-lived. In contrast, the color states do not involve any excited states and are thus immune to spontaneous decay: they are long-lived. Consequently, as long as the system starts and stays in the color manifold, there is no spontaneous decay and the evolution is purely Hamiltonian. However, the kinetic and off-resonance terms might couple bright and color manifolds. As long as these couplings are small compared to the manifold energy separation, the adiabatic approximation holds and the evolution is constrained to the color manifold.  

\textbf{Preparation of the initial color state:} Since the initial internal state $|9/2\rangle_g$ is only projected in the two color states $\ket{R}$ and $\ket{G}$ (see Methods), the system can be adiabatically mapped into the corresponding color subspace. This is done by conveniently ramping the 2-tripod beam intensities and frequencies to prepare the desired initial state. To infer the fidelity of the state preparation, the 2-tripod beams are abruptly switched off and a time-of-flight fluorescence image is recorded after a 9$\,{\rm ms}$ free fall. 
Thanks to the topology of the 2-tripod scheme, coherent photon redistribution among the beams entangles velocity and internal degrees of freedom. For atoms at rest and the laser beam orientation depicted in \fig{fig1}a, the corresponding velocity ${\bf v}_{m_F}$ of the state $\ket{m_F}_g$ are ${\bf v}_{1/2}=4v_r\hat{\textbf{y}}$, ${\bf v}_{3/2}=v_r(-\hat{\textbf{x}} +3\hat{\textbf{y}})$, ${\bf v}_{5/2}=2v_r\hat{\textbf{y}}$, ${\bf v}_{7/2}=v_r(-\hat{\textbf{x}}+\hat{\textbf{y}})$ and ${\bf v}_{9/2}=0$. Here, $v_r = \hbar k/m\approx 6.6\,$mm/s is the single-photon recoil velocity, $\hbar$ is the reduced Planck's constant, $k$ is the laser-beam wavenumber, and $m$ is the strontium atomic mass. The left panel in \fig{fig1}d shows the fluorescence image obtained when the ramp sequence protocol targets the color state $\ket{G}$. Integrating out each velocity distribution component in the fluorescence image, we obtain the different ground state populations $P_{m_F}$ as shown in the right panel in \fig{fig1}d (see Methods). These measurements are compared to the predicted populations, obtained by numerical integration of the optical Bloch equations that model the actual ramp sequence protocol. 
The fidelity of the prepared state, computed as $1-1/2\sum_{m_F=1/2}^{9/2}|P_{m_F,\textrm{exp}}-P_{m_F,\textrm{th}}|$, is $\mathcal{F}= 0.97(2)$. Finally, using the five measured ground state populations, we can reconstruct the actual color state $\ket{\psi_0}$ produced by the ramp sequence protocol up to an irrelevant global phase factor, see Supplementary Section III. The table \ref{tab:G_state} illustrates the reconstruction for the example shown in \fig{fig1}d, in good agreement with the targeted color state $\ket{G}$.

\begin{table}[]
    \centering
    \begin{tabular}{|c|c|c|c|}
          \hline
         $\ket{S}$ & $\ket{R}$ & $\ket{G}$ & $\ket{B}$\\
         \hline
         $|\braket{\psi_0|S}|^2$ & 0.06(3) & 0.94(3) & 0\\
         $\mathrm{Arg}\{\braket{\psi_0|S}\}/\pi$ & -0.53(6) & 0.0 & - \\
         \hline
    \end{tabular}
    \caption{Initial color state prepared $\ket{\psi_0}$ after the ramp sequence protocol. The targeted initial color state is $\ket{G}$.
    }
    \label{tab:G_state}
\end{table}

\begin{figure}
    \centering
    \includegraphics[width=0.49\textwidth]{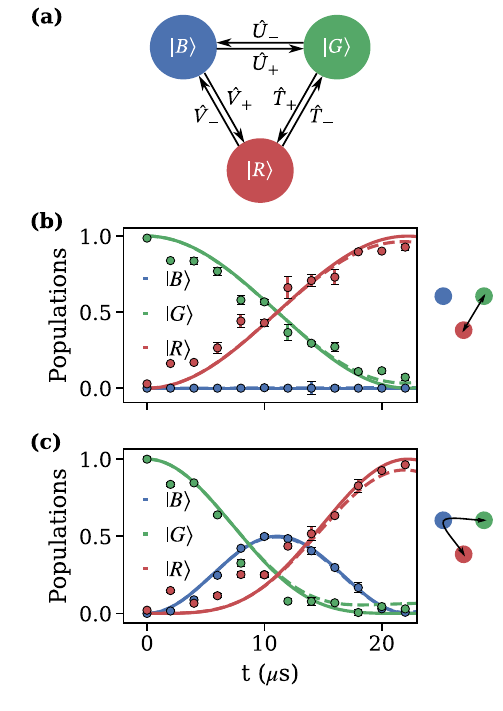}
    \caption{\textbf{Cartan-Weyl transformation}. \textbf{(a)} Schematic of the Cartan-Weyl SU(3) ladder operators structure leading to the all-state connectivity. \textbf{(b)} Time evolution of the populations of states $\ket{R}$ (red disks), $\ket{G}$ (green disks) and $\ket{B}$ (blue disks) for the Hamiltonian $\hat{H}_{RG}$ defined in \eq{CW_H}. With an initial state as $\ket{G}$, we observe a direct $\ket{G}\rightarrow\ket{R}$ population transfer. The colored solid lines are the theoretical predictions for an atom at rest and the colored-dashed curves correspond to theoretical predictions taking into account the momentum dispersion. \textbf{(c)} Same as (b) but for Hamiltonian $\hat{H}_{RGB}$. We observe an indirect $\ket{G}\rightarrow\ket{R}$ population transfer mediated by $\ket{B}$. The practical implementation of the transformations is detailed in Methods. }
    \label{fig:fig2}
\end{figure}

\textbf{SU(3) adiabatic population transfer: } To infer the all-state connectivity of the SU(3) structure, see \fig{fig2}a, we initialize the ultracold gas at rest in the state $\ket{G}$, and we implement two geometric transformations through the action of two specific traceless coupling Hamiltonians. Using conveniently chosen laser detunings (see Methods), we 
generate:
\begin{eqnarray}
\label{eq:CW_H}
&&\hat{H}_{RG} = \hbar\Delta_0 \, (\hat{T}_+ + \hat{T}_-) \nonumber \\
&&\hat{H}_{RGB} =  \sqrt{2} \, \hbar\Delta_0 \, (\hat{V}_++\hat{V}_-+\hat{U}_++\hat{U}_-),
\end{eqnarray} 
where $\Delta_0 = 2\pi\times 11.25\,\textrm{kHz}$ and $\hat{T}_\pm$, $\hat{V}_\pm$ and $\hat{U}_\pm$ are the ladder operators of the Cartan-Weyl basis, see Supplementary Section V. We have $\hat{T}_+\ket{G} = \ket{R}$ (and similar relations for the other operators and states, see \fig{fig2}a). 

When $\hat{H}_{RG}$ is applied to the system, we observe a direct $\ket{G}\rightarrow\ket{R}$ population transfer while $
\ket{B}$ remains spectator, see \fig{fig2}b. In contrast, we observe an indirect $\ket{G}\rightarrow\ket{R}$ population transfer mediated by $
\ket{B}$ when $\hat{H}_{RGB}$ is applied, see \fig{fig2}c. Both behaviors are expected from Eq.\eqref{eq:CW_H} and are in good agreement with a theoretical prediction for an atom at rest (Plain curves). 

The two unitary operations, depicted in \fig{fig2}b and \fig{fig2}c, are generators of a universal qutrit gate \cite{PhysRevApplied.19.034089}. Also, couplings in three-level systems have been proposed for braiding operations \cite{PhysRevA.109.022431}, and atomic interferometers \cite{buckle1986atomic}, and reported in the form of a closed-contour interaction to realize phase-sensitive coherent population trapping \cite{barfuss2018phase}, dynamical decoupling \cite{stark2018clock,finkelstein2021continuous}, and quantum batteries \cite{dou2020closed}.

\textbf{Color oscillations: } Now, we consider the dynamics of our ultracold atomic ensemble under a spatially homogeneous and time-independent SU(3) gauge field which reduces to a color-orbit coupling. The corresponding Hamiltonian is experimentally realized using a specific set of laser detunings to cancel the scalar gauge field components (see Methods). It reads:
\begin{equation} \label{eq:CO_H}
\hat{H}=\frac{\hat{\textbf{p}}^2}{2m} \mathbb{1}_3 -\frac{\hat{\textbf{p}}\cdot\hat{\textbf{A}}}{m},
\end{equation}
where $\hat{\textbf{p}}$ is the momentum operator and $\mathbb{1}_3$ is the identity in the color manifold. 
The nontrivial dynamics is imparted by the color-orbit coupling term $\hat{\textbf{p}}\cdot\hat{\textbf{A}}/m$, where $\hat{\textbf{A}}$ is the artificial SU(3) gauge field represented by $3\times 3$ matrices with entries $\textbf{A}_{IJ}=i\hbar\langle I|\boldsymbol{\nabla} J\rangle$ where $\ket{I}$ and $\ket{J}$ are any of the color states \cite{dalibard2011colloquium}. Since the gauge field is space-independent, it commutes with $\hat{{\bf p}}$ and $[\hat{H},\hat{{\bf p}}]=0$: The momentum operator $\hat{{\bf p}}$ is thus a constant of motion. We stress that the gauge-field components do not commute within each other, meaning that the gauge field is non-Abelian. This is of crucial importance for understanding the dynamics of the atoms. Indeed, using the Heisenberg picture, the atoms are subject to a non-inertial ($\hat{\textbf{p}}$-dependent) non-Abelian force operator \cite{hasan2022wavepacket}
\begin{equation} \label{eq:Force}
\hat{\textbf{F}}=m\frac{d\hat{\textbf{v}}}{dt}=\frac{im}{\hbar}\left[\hat{H},\hat{\textbf{v}}\right]=\frac{i}{m\hbar}\, \hat{\textbf{p}}\times\left(\hat{\textbf{A}}\times\hat{\textbf{A}}\right),
\end{equation} 
where $\hat{\textbf{v}}=(\hat{\textbf{p}}-\hat{\textbf{A}})/m$ is the velocity operator. This force operator, a generalization of the Lorentz force at higher symmetries, would simply vanish for an Abelian gauge field. The cross product $i \hat{\textbf{A}}\times\hat{\textbf{A}}/\hbar$ corresponds to the non-Abelian field curvature, a generalization of the magnetic field strength \cite{dalibard2011colloquium}. 

\begin{figure}[htpb]
\includegraphics[width=0.45\textwidth]{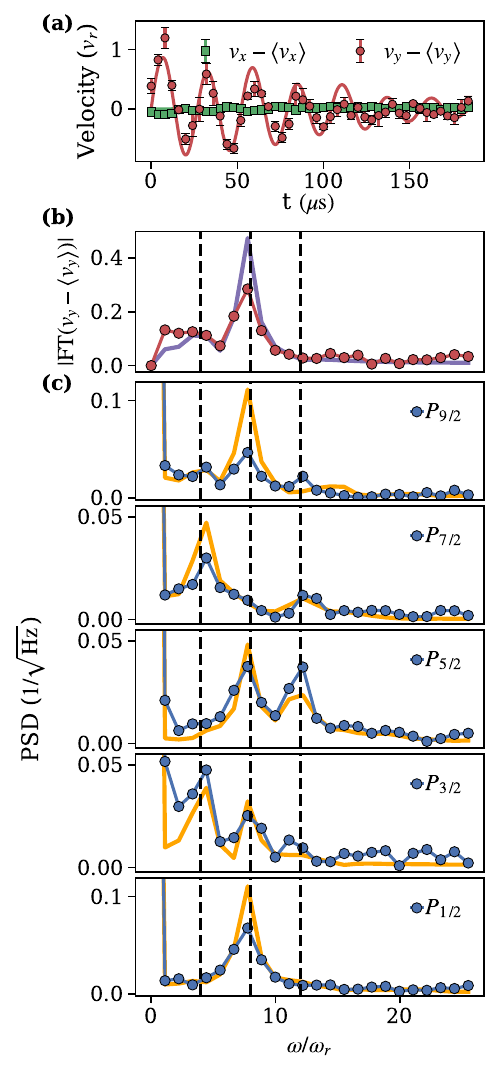}
\caption{\textbf{Color-orbit dynamics}. \textbf{(a)} Temporal evolution of the atomic velocity components $v_x$ (green circles) and $v_y$ (red squares) around their mean values $\langle v_x\rangle= -0.14(2)v_r$ and $\langle v_y\rangle= 1.91(11)v_r$. The time origin corresponds to the end of the adiabatic ramp sequence transferring bare state $\ket{9/2}_g$ to a color state and the start of the evolution in the SU(3) gauge field. The plain curves are the theory prediction considering the finite momentum spread of gas. The (conserved) mean initial momentum of the gas is ${\bf p}=8 \hbar k \,\hat{{\bf x}}$. \textbf{(b)} Fourier Transformation (FT) amplitude of $v_y-\langle v_y\rangle$ obtained by Fourier transform of the time signal in (a). \textbf{(c)} Bare state populations spectral density (PSD) as indicated on each panel. 
In all panels, the plain curves are numerical predictions using \eq{CW_H} and the finite momentum dispersion of the gas, see methods. The dashed black vertical lines indicate the predicted three Bohr oscillation frequencies $4\omega_r$, $8\omega_r$ and $12\omega_r$ for an atom at rest, see methods.}
\label{fig:fig3} 
\end{figure} 

Thanks to the entanglement between spin and velocity woven by photon transfer in the 2-tripod, the velocity can be indirectly extracted from the Zeeman states populations $P_{m_F}$. We have indeed  
\begin{equation} \label{eq:Mv}
{\bf v}(t) = \langle {\hat{\textbf{v}}}(t)\rangle=\sum_{m_F} P_{m_F}(t) \, {\bf v}_{m_F}.
\end{equation} 
In \fig{fig3}a, we show a typical example of the evolution of the ultracold gas mean velocity as a function of the interaction time $t$ in the gauge field. The bare state populations $P_{m_F}(t)$ are extracted from the time-of-flight images, see for example \fig{fig1}d. The key feature of the velocity evolution in \fig{fig3}a is its oscillatory behavior. Its damping, with a characteristic time of $\sim 80\,\mu$s, is just a detrimental effect due to the finite momentum spread of the gas (quantum and thermal) \cite{Hasan2022}. The second important feature is that the oscillatory motion takes place along the $y$-axis. This can be easily understood 
by noting that $\langle \hat{\textbf{v}}(t)\rangle=[\textbf{p}-\langle \hat{\textbf{A}}(t)\rangle]/m$, where ${\bf p}=8\hbar k \, \hat{{\bf x}}$ is the initial momentum of the atomic ensemble in the laser reference frame (see Methods). Hence, the dynamics of the mean velocity is driven by $\langle \hat{\textbf{A}}(t)\rangle$, which in turn oscillates in a direction orthogonal to $\textbf{p}$ as expected from the expression of the force given by \eq{Force}, indicating the Hall-like nature of the color-orbit coupling. Starting from \eq{CO_H}, the full dynamical behavior of the mean velocity is well recovered by a numerical integration of Schr\"odinger equation in the color manifold. The resulting mean bare state populations are computed by taking a random sampling of the finite momentum spread of the atomic gas (see Methods). 

Spin-1/2 systems subject to an SU(2) spin-orbit coupling also develop velocity temporal oscillations. They are associated with the coherent transfer of the atomic wave-packet between the two energy branches of the system \cite{PhysRevB.74.172305,hasan2022wavepacket}. The oscillation frequency is just the Bohr frequency relating to these two states. For SU(3) systems subject to color-orbit coupling, there are three energy branches and three Bohr frequencies are thus involved in general. With our choice of parameters, they read $4\omega_r$, $8\omega_r$, and $12\omega_r$ (see Methods). Here, $\omega_r = kv_r/2\approx 2\pi\times4.8\,$kHz is the recoil frequency. Unexpectedly, as seen in \fig{fig3}b, the amplitude spectrum of the $v_y$ velocity component lacks the contribution from the highest Bohr frequency. However, as seen in \fig{fig3}c, the spectral density of the bare states populations do show contributions from the three expected Bohr frequencies. Overall, the evolution of the internal state shows a richer dynamical behavior than the external degree of freedom. This result contrasts with two-states SU(2) systems, where only one frequency component is present in internal and external degrees of freedom \cite{hasan2022wavepacket}. 

As seen in \fig{fig3}, our theoretical predictions using \eq{CW_H} capture well the dynamical behavior of the atomic velocity. However, even if their positions match well with the predictions, the amplitude of the resonant peaks in the bare-state spectra are less well reproduced. Since these peaks result from a delicate balance between three interference amplitudes, this mismatch is likely due to systematic errors coming from residual AC-stark shifts or imperfections in state preparations, see Supplementary Section IV for a more systematic study. In contrast, the position of the resonant peaks comes from the energy difference between eigenenergies, a quantity that is more robust to experimental imperfections.  

\vspace{0.5cm}
\noindent {\Large\textbf{Discussion}}

We used a 2-tripod scheme to generate an SU(3) non-Abelian gauge field acting on an ultracold cloud of fermionic strontium atoms. We featured the SU(3) nature by performing geometric Cartan-Weyl transformations where two given states can be directly coupled or indirectly coupled via the third state, thereby revealing the all-state connectivity, a genuine property of SU(3) systems. Then, we studied the oscillatory dynamics of the ultracold atomic gas subject to the non-inertial force imparted by a constant and uniform SU(3) gauge field. The systems exhibit unique dynamical behavior characterized by three Bohr frequencies. Again, this feature is rooted in all-state connectivity.

In our study, the SU(3) non-abelian gauge field is reduced to a color-orbit coupling by canceling the geometrical scalar potential with a proper choice of two-photon detunings within the 2-tripod scheme (see values in Supplementary Table II). It results in a triple degeneracy of the energy branches at zero momentum. Adding a scalar potential generally lifts this degeneracy, which reduces to momentum-separated double degeneracies, see Supplementary Section II. As the all-state connectivity is preserved, the dynamics are still characterized by three Bohr frequencies. However, the topological properties might be different and yet need to be explored. Finally, we note that our system evolves according to a SU(3) unitary evolution operator, like the Pontecorvo-Maki-Nakagawa-Sakata matrix used in flavor oscillation of neutrinos \cite{gonzalez2008phenomenology} and the Cabibbo-Kobayashi-Maskawa matrix describing quark mixing \cite{PhysRevLett.10.531,kobayashi1973cp,dattoli2007quark}. Hence, our system could be an ideal platform to explore these high-energy physics systems, as parameters such as coupling strengths (Rabi frequencies), frequency mismatches (detunings), and laser-beam orientation allow for precise control of the matrices entries. It shall be noted, however, that the orientation of the lasers and the nature of the couplings do not allow for general unitary evolutions. For instance, the unitary matrix elements are real, which prevents us to mimic the CP-violating phase in neutrino oscillation\cite{gonzalez2008phenomenology}, in contrast for example to a proposed staggered multi-spin model in lattice \cite{lan2011tunable}.

\vspace{0.5cm}
\noindent {\Large\textbf{Methods}}

\noindent \textbf{Ultracold Sample Preparation:} We prepare a quantum degenerate ultracold sample of $^{87}\text{Sr}$ atoms using, as an initial step, standard laser cooling techniques. Then, $2.5(1) \times 10^6$ atoms at a temperature of $6\ \mu$K are loaded into a far-off-resonant optical-dipole trap formed by crossing two 1064 nm focused beams ($1/e^2$-waist: $60\,\mu$m with a power of 4.5 W each). We perform a partial optical pumping of the ground state Zeeman sublevels such that half the atoms are in the $m_F=9/2$ state and the other half are distributed over the negative Zeeman sublevels $m_F\leq-1/2$. Next, we perform an evaporative cooling sequence by lowering the power of the dipole-trap beams over $8\,$s and obtain $N=5(1)\times10^4$ atoms in the $m_F=9/2$ Zeeman sublevel at a temperature $T=55(4) \,  \textrm{nK} = 0.26(2) \, T_F$, where $T_F$ is the Fermi temperature of the gas. As $T_F/T_r\sim 0.9$, where 
$T_r$ is the single photon recoil temperature, the bare state population are resolved in the time-of-flight images as shown in \fig{fig1}d. Finally, we switch off the dipole trap and begin the state preparation sequence using the 2-tripod laser scheme after the magnetic-field bias of 67 G is turned on.\\

\noindent \textbf{2-tripod laser scheme:} The 2-tripod laser scheme is shown in Fig. \ref{fig:fig1}c. A laser beam from an injection-locked diode laser is split into three beams using non-polarizing beam splitters. Each beam passes through an acousto-optic modulator (AOM) and an electro-optic modulator (EOM). The EOMs are phase modulated to create sidebands at a frequency of 14.12 MHz corresponding to the Zeeman splitting between the excited states $\ket{3/2}_e$ and $\ket{7/2}_e$. While the carriers are resonantly coupled to one excited state of the 2-tripod level scheme, one sideband on each beam resonantly couples the other excited state. For the $\sigma^+$ and $\sigma^-$ transitions, the carrier couples the excited state $\ket{7/2}_e$ and the -1 sideband couples the excited state $\ket{3/2}_e$. For the $\pi$ transitions, the carrier couples the $\ket{3/2}_e$ and the +1 sideband couples the excited state $\ket{7/2}_e$. This choice is made according to the Clebsch-Gordan coefficients of the transitions that minimize the EOMs modulation depth and reduce the power on the unwanted sidebands. The remaining sidebands do not play a decisive role as they are either far off-resonant to atomic transitions, or connect empty states located outside the 2-tripod level scheme. 
Accounting for the Zeeman shift of 12.4 kHz between successive ground state sublevels, the AOMs are used to set the carrier frequencies on resonance with the relevant transitions of the 2-tripod scheme. The Rabi frequencies and  detunings of all the transitions of the 2-tripod scheme are controlled by tuning the powers and frequencies of the RF signals sent to the AOMs and EOMs, see Supplementary Section I for more details.  

The experiments are performed within the adiabatic limit, meaning that the state evolution is constrained within the color manifold. This happens when the Doppler effect $k\bar{v}$ ($k$ is the laser wavevector, $\bar{v}$ the gas $rms$ velocity) and all laser detunings $\delta^{l,r}_{1,2,3}$ are much smaller than the laser Rabi frequency $\Omega$. In the experiment, $k\bar{v}= 2\pi\times 6.5(3)\,$kHz and $\Omega= 2\pi\times 228(10)\,$kHz, while the detunings are kept below $2\pi\times40\,$kHz.

\noindent \textbf{Dark states and Color states:} The interaction Hamiltonian of the 2-tripod scheme hosts a 3-fold degenerate manifold that is uncoupled to the laser fields. We consider the following dark-state basis in this degenerate manifold:
\begin{align}
    \label{eq:DlD0Dr_basis}
	\ket{D_l} =& \sin \varphi_l \, e^{4iky}\ket{1} - \cos \varphi_l \, e^{ik(-x+3y)}\ket{2} \nonumber \\
	\ket{D_0} =& \frac{1}{\alpha_0} \big[\cot \vartheta_l \, (\cos \varphi_l \, e^{4iky}\ket{1} + \sin \varphi_l \, e^{ik(-x+3y)}\ket{2}) - \nonumber \\ & e^{2iky}\ket{3} + 
	\cot \vartheta_r \, ( \sin \varphi_r \, e^{ik(-x+y)}\ket{4} + \cos \varphi_r \, \ket{5})\big] \nonumber \\
	\ket{D_r} =& \sin \varphi_r \ket{5} -\cos \varphi_r \, e^{ik(-x+y)}\ket{4}.
\end{align}
Here, $\tan\varphi_a = |\Omega_2^a/\Omega_1^a|$, $\cos\vartheta_a =  \frac{|\Omega^a_3|}{\sqrt{|\Omega^a_1|^2+|\Omega^a_2|^2+|\Omega^a_3|^2}}$ and $\alpha_0=\sqrt{1+\cot^2{\vartheta_l}+\cot^2{\vartheta_r}}$, where $a=l,r$ refers to the left ($l$) or right ($r$) tripod. For equal Rabi frequencies, we have $\varphi_a=\pi/4$, $\cos\vartheta_a = \sqrt{3}/3$ and $\alpha_0=\sqrt{2}$. The expressions in \eq{DlD0Dr_basis} are given for laser beams at resonance. A more general expression is given in Supplementary Section II, and the modulation depths of the EOMs for equal Rabi frequencies are reported in Supplementary Table I.

The color states are chosen to facilitate the generation of the coupling term depicted in \eq{CW_H}. They read:
\begin{align}
    \label{color-basis}
    \ket{R} &= \frac{\ket{D_0}-e^{i\pi/4}\ket{D_r}}{\sqrt{2}}\nonumber \\
	\ket{G} &= \frac{\ket{D_0}+e^{i\pi/4}\ket{D_r}}{\sqrt{2}}\\
    \ket{B} &= \ket{D_l}. \nonumber 
\end{align}

\noindent\textbf{Initial state preparation and Cartan-Weyl transformations:} We map the atoms from the stretched state $\ket{9/2}_g$ into the dark-state subspace by adiabatically ramping the Rabi frequencies of the 2-tripod lasers. For the color-orbit coupling experiment, we first abruptly turn on the $\sigma^+$ and $\pi$ beams along with their sidebands. This operation does not affect the atomic state that remains $\ket{9/2}_g$. Then, we ramp the $\sigma^-$ laser such that $\Omega_3^l(t) = \Omega_1^r(t) = \Omega\tan \left(\frac{ \pi t}{4 t_{\text{\textrm{ramp}}}}\right)$, where $t_{\text{\textrm{ramp}}}=18\, \mu$s is the duration of the ramp sequence, see Supplementary Section III for more details. 

The initial state $\ket{G}$ for Cartan-Weyl transformation is prepared using the same Rabi frequencies ramps as above but with additional detunings of $\delta^r_3 = 2\delta^r_2 = 2\pi \times 50$ kHz. 
Once the ramps are completed, the detunings are changed to $\delta_1^r = \delta_2^r = 2\pi \times 30$ kHz and $\delta_3^r=\delta_i^l=0$ for $i=\{1,2,3\}$ to achieve $\hat{H}_{RG}$. Similarly, detunings $\delta_1^l = -\delta_2^l = 2\pi \times 45$ kHz and $\delta_3^l = \delta_i^r = 0$ for $i=\{1,2,3\}$ to implement $\hat{H}_{RGB}$. The resulting traceless Hamiltonians read (see Supplementary Section II)
\begin{align}
    \hat{H}_{RG} =  \frac{\hbar \delta_1^r}{8} \left(3\hat{\lambda}_1 - \frac{5\hat{\lambda}_8}{\sqrt{3}}\right) \\
   \hat{H}_{RGB}= -\frac{\hbar \delta_1^l}{2\sqrt{2}}\left(\hat{\lambda}_4+\hat{\lambda}_6\right),
\end{align}
corresponding to the Hamitonians in \eq{CW_H}, with an extra diagonal term proportional to $\lambda_8$ that does not play any role as the evolution given by $\hat{H}_{RG}$ is restricted to the $\{\ket{R}, \ket{G}\}$ subspace. The total evolution time is of $22\,\mu$s (see Fig. \ref{fig:fig2}), which is short enough to consider that the gas remains at rest.

\noindent \textbf{Color-orbit Hamiltonian:} Restricting the Schr\"odinger equation to the dark-state subspace (adiabatic approximation), we get the following effective Hamiltonian:
\begin{equation}
    \label{HamTotal}
    \hat{H} = \frac{\hat{\textbf{p}}^2}{2m}-\frac{\hat{\textbf{p}}\cdot\hat{\textbf{A}}}{m} + \hat{W} + \hat{\Delta},
\end{equation}
where $\boldsymbol{A}_{IJ}=i\hbar\langle I|\boldsymbol{\nabla} J\rangle$ is the artificial gauge field, $W_{IJ} = -\frac{\hbar^2}{2 m} \langle \boldsymbol{\nabla} I| \boldsymbol{\nabla} J\rangle$ is the geometric scalar potential, and $\Delta_{IJ} = -i\hbar\langle I|\partial_t| J\rangle$ ($\{I,J\}=\{R,G,B\}$) is another scalar term that arises due to temporal variations of the dark-states parameters such as the Rabi frequencies and the detunings. As we vary only the detuning, we will call this operator the detuning matrix. See Supplementary Section II for a complete expression of these matrices.

In the following, we cancel the scalar potential $\hat{W}$ by an appropriate set of detunings such as $\hat{\Delta} = -\hat{W}$, leading to the color-orbit coupling Hamiltonian of \eq{CO_H}. In the atomic reference frame, this condition is equivalent to the 2-tripod two-photons transitions being at resonance, where we take into consideration the recoil frequency shifts associated with the photon redistribution. Here, the detunings set is not unique and it is chosen to minimize the single-photon detunings.

Using the color-state basis, and expanding over the Gell-Mann matrices, we get the gauge field in matrix form, $\hat{\textbf{A}} = \hbar k \big(\boldsymbol{a}_0 \, \hat{\mathbbm{1}}_3 + \sum_{=1}^{8} \boldsymbol{a}_i \, \hat{\lambda}_i\big)$. Using $\textrm{Tr}(\hat{\lambda}_i\hat{\lambda}_j)=2\delta_{ij}$, we find $\boldsymbol{a}_0=\textrm{Tr}\hat{\textbf{A}}/(3 \hbar k)$ and $\boldsymbol{a}_i = \textrm{Tr}(\hat{\lambda}_i\hat{\textbf{A}})/(2\hbar k)$. Within our experimental framework, namely with equal Rabi frequencies and orientation of the laser beams as depicted in Fig. 1c, we have: 
\begin{eqnarray}
  &&\boldsymbol{a}_0 = \frac{1}{12} (5\hat{\textbf{x}}-24\hat{\textbf{y}}) \quad \boldsymbol{a}_1 = - \frac{1}{8} (\hat{\textbf{x}}+6\hat{\textbf{y}}) \nonumber\\
  &&\boldsymbol{a}_2 = \boldsymbol{a}_3 = \frac{\sqrt{2}}{8} (\hat{\textbf{x}}-\hat{\textbf{y}}) \quad
  \boldsymbol{a}_4 = \boldsymbol{a}_6 = -\frac{\sqrt{2}}{8} (\hat{\textbf{x}}+\hat{\textbf{y}})\\
  &&\boldsymbol{a}_5 = \boldsymbol{a}_7 = 0 \quad \boldsymbol{a}_8=-\frac{\sqrt{3}}{24} (\hat{\textbf{x}}-18\hat{\textbf{y}}). \nonumber
\end{eqnarray}
Trivially, $-\hat{\textbf{p}}\cdot\hat{\textbf{A}}/m = \mu_0 \hat{\mathbbm{1}} + \sum_{i=1}^{8} \mu_i \, \hat{\lambda}_i$ where $\mu_i = -v_r \,\textbf{p}\cdot\boldsymbol{a}_i$ ($j=0, ..., 8$). The term $\mu_0\hat{\mathbbm{1}}$ being irrelevant for the dynamics (global phase factor), we can safely ignore it.

Alternatively, using the Cartan-Weyl basis of the $\mathfrak{su}$(3) Lie algebra, namely $\hat{T}_\pm = (\hat{\lambda}_1 \pm i \hat{\lambda}_2)/2$, $\hat{I}_3=\hat{\lambda}_3/2$, $\hat{V}_\pm = (\hat{\lambda}_4 \pm i \hat{\lambda}_5)/2$, $\hat{U}_\pm = (\hat{\lambda}_6 \pm i \hat{\lambda}_7)/2$ and $\hat{Y}=\hat{\lambda}_8/\sqrt{3}$, we have the equivalent decomposition $-\hat{\textbf{p}}\cdot\hat{\textbf{A}}/m= \boldsymbol{\beta}_0 \cdot \hat{\boldsymbol{\tau}}_0 + \boldsymbol{\beta}\cdot\hat{\boldsymbol{\tau}}$, where $\hat{\boldsymbol{\tau}}_0=(\hat{I}_3,\hat{Y})$ regroups the diagonal traceless matrices and $\boldsymbol{\tau}=(\hat{T}_+, \hat{T}_-,\hat{V}_+,\hat{V}_-,\hat{U}_+,\hat{U}_-)$ regroups the Cartan-Weyl ladder operators. We find
\begin{eqnarray} \label{eq:HRGB}
&&\boldsymbol{\beta}_0=(2\mu_3,\sqrt{3}\mu_8) \\
&&\boldsymbol{\beta}=(\mu_1-i\mu_2,\mu_1+i\mu_2, \mu_4,\mu_4,\mu_6,\mu_6),
\end{eqnarray}
and with the initial mean momentum state ${\bf p}=8\hbar k \, \hat{{\bf x}}$, we have:
\begin{eqnarray}
-\hat{\textbf{p}}\cdot\hat{\textbf{A}}/m &=&\hbar\Delta \, [\sqrt{3} \, (u^*\hat{T}_++u\hat{T}_-) - \sqrt{2} \, \hat{I}_3 + \hat{Y}_3/\sqrt{3} \nonumber \\
&+& \sqrt{2} \, (\hat{V}_+ + \hat{V}_- +\hat{U}_+ + \hat{U}_-)],
\end{eqnarray}
where $u=(1+i\sqrt{2})/\sqrt{3}$ and $\Delta = kv_r = 2\omega_r$ is twice the recoil angular frequency $\omega_r = kv_r/2$.\\

\noindent\textbf{SU(3) spectral properties:} Following \cite{kusnezov1995exact}, the secular equation of the SU(3) traceless Hamiltonian matrix $\hat{H} = \sum_{i=1}^{8}\alpha_i \hat{\lambda}_i \equiv \boldsymbol{\alpha}\cdot\hat{\boldsymbol{\lambda}}$
is given by
\begin{equation} \label{eq:Se}
\textrm{Det}(\hat{H}-E\, \hat{\mathbbm{1}}_3)=-E^3+\frac{C_2}{2}E+\frac{C_3}{3}=0. 
\end{equation} 
Writing its eigenenergies $E_i=(2\alpha/\sqrt{3}) \, \epsilon_i$ ($i=1,2,3$), with $\alpha = \sqrt{\boldsymbol{\alpha}^2}$ (note that $\boldsymbol{\alpha}$ is real), we have 
\begin{eqnarray} \label{eq:Ev}
\epsilon_1 &=& \sin\left(\gamma+\frac{\pi}{3}\right)\noindent \nonumber\\
\epsilon_2 &=&\sin\left(\gamma-\frac{\pi}{3}\right)\noindent\\
\epsilon_3 &=& -\sin\gamma, \nonumber
\end{eqnarray}
where $\gamma\in[-\pi/3,\pi/3]$. Using the condition $\textrm{Tr}\hat{H}=0$, one can easily show that $C_2=\textrm{Tr}(\hat{H}^2)=2\alpha^2$ and 
$C_3=\textrm{Tr}(\hat{H}^3)=(2\sqrt{3}/3)\alpha^3\sin3\gamma$. They are the two Casimir invariants of SU(3). 

From Eqs \ref{eq:Se}, we note that the energy gaps between consecutive eigenenergies are equal when $C_3=0$. Noting that $\sin3\gamma = \sqrt{6} C_3 C_2^{-3/2}$, this happens when $\gamma =0, \pm \pi/3$. In this case, the symmetry of the system reduces to SU(2) and the Hamiltonian takes the simpler form $\hat{H}=\boldsymbol{\beta}\cdot\hat{\mathbf{J}}$, where $\hat{\mathbf{J}}$ is the angular momentum operator of a spin one.
The secular equation is now $-E^3+C_2E=0$, with $C_2=\textrm{Tr}(\hat{H}^2)=2\beta^2$, and the eigenenergies are evenly separated by $\sqrt{C_2}=\sqrt{2}\beta$. Since only one raising and one lowering operators exist, the so-called $\hat{J}_+$ and $\hat{J}_-$ operators, the temporal dynamics is characterized by a single frequency scale, namely $\sqrt{C_2}/\hbar$, and can involve at most two characteristic Bohr frequencies, namely $\sqrt{C_2}/\hbar$ and $2\sqrt{C_2}/\hbar$.

When $C_3 \neq 0$, the two consecutive energy gaps are different, and the color-orbit temporal dynamics involve up to three different Bohr frequencies (one being the sum of the two others). This is a genuine signature that the system symmetry is truly SU(3) and not reducible to SU(2).
For the spin-orbit coupling Hamiltonian, \eq{CO_H}, we get
\begin{eqnarray} \label{eq:CI}
C_2 &=& v_r^2 \, \frac{7p_x^2+ 114p_y^2}{24}\noindent\\
C_3 &=& v_r^3 \, \frac{p_x (10p_x^2+9p_y^2)}{288},
\end{eqnarray}
and the gauge field is truly SU(3) as long as $p_x \neq 0$.

With our initial momentum state ${\bf p}=8\hbar k \, \hat{{\bf x}}$, we find 
$\gamma \approx 0.19$ rd and the Bohr frequencies read $4\omega_r$, $8\omega_r$ and $12\omega_r$.

\noindent\textbf{Momentum boost:} The dynamics of the ultracold gas in the color-orbit coupling Hamiltonian \eq{CO_H} is observed by adding an initial momentum kick to explore an area of the dispersion relation where the eigenenergy separations are much larger than the mean kinetic energy $\sim\hbar\omega_r$. By doing so, the temporal oscillations of the velocity can be observed at a time shorter than the damping time. In addition, the energy separations should be much smaller that the Rabi frequencies $\sim54\omega_r$ used in this protocol. In the experiment, the momentum boost is defined with respect to the laser beams reference frame. Because of the Doppler effect, a mean momentum $\textbf{p}=p_x \hat{{\bf x}} + p_y \hat{{\bf y}}$ in the laser beam frame corresponds to the following set of detunings in the atomic frame:
\begin{align}
    \label{det_momentum}
    \delta_1^r = \delta_3^l &= -k p_y/m \nonumber\\
    \delta_2^r = \delta_2^r &= -k p_x/m \\
    \delta_3^r = \delta_1^l &= k p_y/m. \nonumber
\end{align}
This means that the detuning matrix reads $\hat{\Delta} = -(p_x \hat{A}_x + p_y \hat{A}_y)/m $, see Supplementary Section II. The data shown in Fig. \ref{fig:fig3} correspond to a momentum boost of $\textbf{p}=8 \hbar k \, \hat{{\bf x}}$.

\noindent\textbf{Image analysis:}
The population of bare states is obtained by fitting time-of-flight images with the fitting function
$F = \Sigma_{i=1}^{5} G(A_i, x_{0i}, y_{0i}, \sigma_{xi}, \sigma_{yi}, \theta_i) + \text{offset}$. Here, $G(A_i, x_{0i}, y_{0i}, \sigma_{xi}, \sigma_{yi}, \theta_i)$ is a 2D-Gaussian function with an amplitude $A_i$, centered at coordinates $(x_{0i}, y_{0i})$, with rms widths $\sigma_{xi}$ and $\sigma_{yi}$ oriented along an axis making an angle $\theta_i$ with the x-axis. To obtain a reliable fit, we first crop the image around the cloud corresponding to $p=0$, which includes atoms in the ground state $m_F=9/2$. We fit that distribution with a 2D-Gaussian function to extract the center, the widths, and the offset.  Using these parameters, we constrain the centers of the other 2D-Gaussian fits such that they are at the expected mean momentum of the corresponding bare state. In addition, the widths of the Gaussian fit are restricted to be within 10\% of the widths obtained from the single 2D-Gaussian fit. The total area under the fits is proportional to the total number of atoms. Of these, $50\%$ of the atoms are spectators in negative $m_F$ ground states, and their contribution is removed from the $\hat{\textbf{p}}=0$ Gaussian. We find the area under each 2D Gaussian and normalize it to get the population of each $j$-ground state involved in the 2-tripod, that is $P_j = A_j \sigma_{xj}\sigma_{yj}/N$ where $N=\frac{1}{2}\Sigma_{i=1}^{5}A_i \sigma_{xi}\sigma_{yi}$.

\noindent\textbf{Numerical integration:} The numerical curves in Fig. \ref{fig:fig3} are calculated by solving the Schr\"odinger equation with the Hamiltonian of \eq{CO_H} in the color manifold with appropriate initial state preparation as discussed in the main text and the previous subsections. The final state is converted into bare state population by using the relationships in \eq{DlD0Dr_basis}. The momentum spread due to the gas temperature and degeneracy is taken into account by solving the Schr\"odinger equation for 1000 momentum values randomly sampled from the experimental momentum distribution of the ultracold gas.\\

\vspace{0.5cm}
\noindent {\Large\textbf{Data Availability}}

The data generated in this study have been deposited in Dataverse under the accession code \url{https://doi.org/10.21979/N9/HECOKO}.

\vspace{0.5cm}

\bibliographystyle{plain}

\begin{thebibliography}{10}

\bibitem{aidelsburger2015measuring}
Monika Aidelsburger, Michael Lohse, Christian Schweizer, Marcos Atala, Julio~T Barreiro, Sylvain Nascimb{\`e}ne, NR~Cooper, Immanuel Bloch, and Nathan Goldman.
\newblock Measuring the chern number of hofstadter bands with ultracold bosonic atoms.
\newblock {\em Nat. Phys.}, 11(2):162--166, 2015.

\bibitem{banerjee2013}
D.~Banerjee, M.~B\"ogli, M.~Dalmonte, E.~Rico, P.~Stebler, U.-J. Wiese, and P.~Zoller.
\newblock Atomic quantum simulation of $\mathbf{U}(n)$ and $\mathrm{SU}(n)$ non-abelian lattice gauge theories.
\newblock {\em Phys. Rev. Lett.}, 110:125303, 2013.

\bibitem{barfuss2018phase}
Arne Barfuss, Johannes K\"{o}lbl, Lucas Thiel, Jean Teissier, Mark Kasperczyk, and Patrick Maletinsky.
\newblock Phase-controlled coherent dynamics of a single spin under closed-contour interaction.
\newblock {\em Nat. Phys.}, 14(11):1087--1091, 2018.

\bibitem{barfussphasecontrolled2018}
Arne Barfuss, Johannes Kölbl, Lucas Thiel, Jean Teissier, Mark Kasperczyk, and Patrick Maletinsky.
\newblock Phase-controlled coherent dynamics of a single spin under closed-contour interaction.
\newblock {\em Nat. Phys.}, 14(11):1087.

\bibitem{beeler2013spin}
Matthew~C Beeler, Ross~A Williams, Karina Jimenez-Garcia, Lindsay~J LeBlanc, Abigail~R Perry, and Ian~B Spielman.
\newblock The spin hall effect in a quantum gas.
\newblock {\em Nature}, 498(7453):201--204, 2013.

\bibitem{PhysRevA.98.043614}
U.~Bornheimer, C.~Miniatura, and B.~Gr\'emaud.
\newblock Su(3) topological insulators in the honeycomb lattice.
\newblock {\em Phys. Rev. A}, 98:043614, Oct 2018.

\bibitem{braibant2011}
Sylvie Braibant, Giorgio Giacomelli, and Maurizio Spurio.
\newblock {\em Particles and Fundamental Interactions: An Introduction to Particle Physics}.
\newblock Springer Science \& Business Media, 2011.

\bibitem{buckle1986atomic}
SJ~Buckle, SM~Barnett, PL~Knight, MA~Lauder, and DT~Pegg.
\newblock Atomic interferometers.
\newblock {\em Opt. Act.: Int. J. Opt.}, 33(9):1129--1140, 1986.

\bibitem{PhysRevLett.10.531}
Nicola Cabibbo.
\newblock Unitary symmetry and leptonic decays.
\newblock {\em Phys. Rev. Lett.}, 10:531--533, Jun 1963.

\bibitem{cooper2019topological}
NR~Cooper, J~Dalibard, and IB~Spielman.
\newblock Topological bands for ultracold atoms.
\newblock {\em Rev. Mod. Phys.}, 91(1):015005, 2019.

\bibitem{PhysRevB.74.172305}
J\'ozsef Cserti and Gyula D\'avid.
\newblock Unified description of zitterbewegung for spintronic, graphene, and superconducting systems.
\newblock {\em Phys. Rev. B}, 74:172305, Nov 2006.

\bibitem{dalibard2011colloquium}
Jean Dalibard, Fabrice Gerbier, Gediminas Juzeli{\=u}nas, and Patrik {\"O}hberg.
\newblock Colloquium: Artificial gauge potentials for neutral atoms.
\newblock {\em Rev. Mod. Phys.}, 83(4):1523, 2011.

\bibitem{dattoli2007quark}
G~Dattoli and K~Zhukovsky.
\newblock Quark mixing in the standard model and space rotations.
\newblock {\em Eur. Phys. J. C}, 52:591--595, 2007.

\bibitem{dirac1927quantum}
Paul Adrien~Maurice Dirac.
\newblock The quantum theory of the emission and absorption of radiation.
\newblock {\em Proc. R. Soc. Lond. A}, 114(767):243--265, 1927.

\bibitem{dou2020closed}
Fu-Quan Dou, Yuan-Jin Wang, and Jian-An Sun.
\newblock Closed-loop three-level charged quantum battery.
\newblock {\em Europhys. Lett.}, 131(4):43001, 2020.

\bibitem{PhysRevLett.105.143902}
Felix Dreisow, Matthias Heinrich, Robert Keil, Andreas T\"unnermann, Stefan Nolte, Stefano Longhi, and Alexander Szameit.
\newblock Classical simulation of relativistic zitterbewegung in photonic lattices.
\newblock {\em Phys. Rev. Lett.}, 105:143902, Sep 2010.

\bibitem{finkelstein2021continuous}
Ran Finkelstein, Ohr Lahad, Itsik Cohen, Omri Davidson, Shai Kiriati, Eilon Poem, and Ofer Firstenberg.
\newblock Continuous protection of a collective state from inhomogeneous dephasing.
\newblock {\em Phys. Rev. X}, 11(1):011008, 2021.

\bibitem{PhysRevLett.129.100501}
Yue Fu, Wenquan Liu, Xiangyu Ye, Ya~Wang, Chengjie Zhang, Chang-Kui Duan, Xing Rong, and Jiangfeng Du.
\newblock Experimental investigation of quantum correlations in a two-qutrit spin system.
\newblock {\em Phys. Rev. Lett.}, 129:100501, Aug 2022.

\bibitem{gerritsma2010quantum}
Rene Gerritsma, Gerhard Kirchmair, Florian Z{\"a}hringer, E~Solano, R~Blatt, and CF~Roos.
\newblock Quantum simulation of the dirac equation.
\newblock {\em Nature}, 463(7277):68--71, 2010.

\bibitem{goldman2014light}
Nathan Goldman, G~Juzeli{\=u}nas, Patrik {\"O}hberg, and Ian~B Spielman.
\newblock Light-induced gauge fields for ultracold atoms.
\newblock {\em Rep. Prog. Phys.}, 77(12):126401, 2014.

\bibitem{gonzalez2008phenomenology}
Maria~Concepci{\'o}n Gonzalez-Garcia and Michele Maltoni.
\newblock Phenomenology with massive neutrinos.
\newblock {\em Phys. Rep.}, 460(1-3):1--129, 2008.

\bibitem{grass2014spiral}
Tobias Gra{\ss}, Ravindra~W Chhajlany, Christine~A Muschik, and Maciej Lewenstein.
\newblock Spiral spin textures of a bosonic mott insulator with su (3) spin-orbit coupling.
\newblock {\em Phys. Rev. B}, 90(19):195127, 2014.

\bibitem{greiner2007}
Walter Greiner, Stefan Schramm, and Eckart Stein.
\newblock {\em Quantum Chromodynamics}.
\newblock CSpringer Berlin, Heidelberg, 2007.

\bibitem{Gross1996}
D.~J. Gross.
\newblock The role of symmetry in fundamental physics.
\newblock {\em Proc. Natl. Acad. Sci. USA}, 93:14256, 1996.

\bibitem{PhysRevB.101.245159}
Mohsen Hafez-Torbati, Jun-Hui Zheng, Bernhard Irsigler, and Walter Hofstetter.
\newblock Interaction-driven topological phase transitions in fermionic su(3) systems.
\newblock {\em Phys. Rev. B}, 101:245159, Jun 2020.

\bibitem{PhysRevA.109.022431}
Zhi-Wei Han, Jia-Hao Liang, Zhao-Xin Fu, Hong-Zhi Liu, Zi-Yuan Chen, Meng Wang, Ze-Rui He, Jia-Yi Huang, Qing-Xian Lv, Kai-Yu Liao, and Yan-Xiong Du.
\newblock Detecting a topological transition of quantum braiding in a threefold-degenerate eigensubspace.
\newblock {\em Phys. Rev. A}, 109:022431, Feb 2024.

\bibitem{Hasan2022}
M.~Hasan, Ch.S. Madasu, K.D. Rathod, C.C. Kwong, and D.~Wilkowski.
\newblock Evolution of an ultracold gas in a non-abelian gauge field: finite temperature effect.
\newblock {\em Quant. Electron.}, 52(6):532--537, jun 2022.

\bibitem{hasan2022wavepacket}
Mehedi Hasan, Chetan~Sriram Madasu, Ketan~D. Rathod, Chang~Chi Kwong, Christian Miniatura, Fr\'ed\'eric Chevy, and David Wilkowski.
\newblock Wave packet dynamics in synthetic non-abelian gauge fields.
\newblock {\em Phys. Rev. Lett.}, 129:130402, Sep 2022.

\bibitem{hu2014u}
Yu-Xin Hu, Christian Miniatura, David Wilkowski, and Beno{\^\i}t Gr{\'e}maud.
\newblock U (3) artificial gauge fields for cold atoms.
\newblock {\em Phys. Rev. A}, 90(2):023601, 2014.

\bibitem{kobayashi1973cp}
Makoto Kobayashi and Toshihide Maskawa.
\newblock Cp-violation in the renormalizable theory of weak interaction.
\newblock {\em Prog. Theor. Phys.}, 49(2):652--657, 1973.

\bibitem{konig2007quantum}
Markus Konig, Steffen Wiedmann, Christoph Brune, Andreas Roth, Hartmut Buhmann, Laurens~W Molenkamp, Xiao-Liang Qi, and Shou-Cheng Zhang.
\newblock Quantum spin hall insulator state in hgte quantum wells.
\newblock {\em Science}, 318(5851):766--770, 2007.

\bibitem{PhysRevA.97.023632}
Doga~Murat Kurkcuoglu and C.~A.~R. S\'a~de Melo.
\newblock Color superfluidity of neutral ultracold fermions in the presence of color-flip and color-orbit fields.
\newblock {\em Phys. Rev. A}, 97:023632, Feb 2018.

\bibitem{kusnezov1995exact}
Dimitri Kusnezov.
\newblock Exact matrix expansions for group elements of su (n).
\newblock {\em J. Math. Phys.}, 36(2):898--906, 1995.

\bibitem{lan2011tunable}
Zhihao Lan, A~Celi, W~Lu, P~{\"O}hberg, and M~Lewenstein.
\newblock Tunable multiple layered dirac cones in optical lattices.
\newblock {\em Phys. Rev. Lett.}, 107(25):253001, 2011.

\bibitem{leblanc2013direct}
Lindsay~J LeBlanc, MC~Beeler, Karina Jimenez-Garcia, Abigail~R Perry, Seiji Sugawa, RA~Williams, and Ian~B Spielman.
\newblock Direct observation of zitterbewegung in a bose--einstein condensate.
\newblock {\em New J. Phys.}, 15(7):073011, 2013.

\bibitem{PhysRevLett.120.180401}
F.~M. Leupold, M.~Malinowski, C.~Zhang, V.~Negnevitsky, A.~Cabello, J.~Alonso, and J.~P. Home.
\newblock Sustained state-independent quantum contextual correlations from a single ion.
\newblock {\em Phys. Rev. Lett.}, 120:180401, May 2018.

\bibitem{PhysRevApplied.19.034089}
Joseph Lindon, Arina Tashchilina, Logan~W. Cooke, and Lindsay~J. LeBlanc.
\newblock Complete unitary qutrit control in ultracold atoms.
\newblock {\em Phys. Rev. Appl.}, 19:034089, Mar 2023.

\bibitem{condmat1010002}
Ipsita Mandal and Atri Bhattacharya.
\newblock Cold atoms in u(3) gauge potentials.
\newblock {\em Condens. Matter}, 1(1), 2016.

\bibitem{PhysRevLett.126.210504}
A.~Morvan, V.~V. Ramasesh, M.~S. Blok, J.~M. Kreikebaum, K.~O'Brien, L.~Chen, B.~K. Mitchell, R.~K. Naik, D.~I. Santiago, and I.~Siddiqi.
\newblock Qutrit randomized benchmarking.
\newblock {\em Phys. Rev. Lett.}, 126:210504, May 2021.

\bibitem{ohlsson2011}
Tommy Ohlsson.
\newblock {\em Relativistic Quantum Physics: From Advanced Quantum Mechanics to Introductory Quantum Field Theory}.
\newblock Cambridge University Press, 2011.

\bibitem{PhysRevA.88.021604}
Chunlei Qu, Chris Hamner, Ming Gong, Chuanwei Zhang, and Peter Engels.
\newblock Observation of zitterbewegung in a spin-orbit-coupled bose-einstein condensate.
\newblock {\em Phys. Rev. A}, 88:021604, Aug 2013.

\bibitem{stark2018clock}
Alexander Stark, Nati Aharon, Alexander Huck, Haitham~AR El-Ella, Alex Retzker, Fedor Jelezko, and Ulrik~L Andersen.
\newblock Clock transition by continuous dynamical decoupling of a three-level system.
\newblock {\em Sci. Rep.}, 8(1):14807, 2018.

\bibitem{t1974magnetic}
Gerardus t~Hooft.
\newblock Magnetic monopoles in unified theories.
\newblock {\em Nucl. Phys. B}, 79(CERN-TH-1876):276--284, 1974.

\bibitem{tagliacozzo2013simulation}
L~Tagliacozzo, A~Celi, P~Orland, MW~Mitchell, and M~Lewenstein.
\newblock Simulation of non-abelian gauge theories with optical lattices.
\newblock {\em Nat. Commun.}, 4(1):2615, 2013.

\bibitem{Wang2020}
Ji-Guo Wang, Yue-Qing Li, and Yu-Fei Dong.
\newblock Lattice configurations in spin-1 bose–einstein condensates with the su(3) spin–orbit coupling*.
\newblock {\em Chin. Phys. B}, 29(10):100304, oct 2020.

\bibitem{yang2015high}
Tao Yang, Kanhaiya Pandey, Mysore~Srinivas Pramod, Frederic Leroux, Chang~Chi Kwong, Elnur Hajiyev, Zhong~Yi Chia, Bess Fang, and David Wilkowski.
\newblock A high flux source of cold strontium atoms.
\newblock {\em Eur. Phys. J. D.}, 69(10):226, 2015.

\bibitem{PhysRevA.103.043302}
Man~Hon Yau and C.~A.~R. S\'a~de Melo.
\newblock Eigenspectrum, chern numbers and phase diagrams of ultracold color-orbit-coupled su(3) fermions in optical lattices.
\newblock {\em Phys. Rev. A}, 103:043302, Apr 2021.

\bibitem{PhysRevA.106.053313}
Man~Hon Yau and C.~A.~R. S\'a~de Melo.
\newblock Quantum hall response of su(3) fermions.
\newblock {\em Phys. Rev. A}, 106:053313, Nov 2022.

\bibitem{zhai2015degenerate}
Hui Zhai.
\newblock Degenerate quantum gases with spin--orbit coupling: a review.
\newblock {\em Rep. Prog. Phys.}, 78(2):026001, 2015.

\bibitem{zhang2019recent}
Shanchao Zhang and Gyu-Boong Jo.
\newblock Recent advances in spin-orbit coupled quantum gases.
\newblock {\em JPCS}, 128:75--86, 2019.

\bibitem{zhang2014}
X.~Zhang, M.~Bishof, S.~L. Bromley, C.~V. Kraus, M.~S. Safronova, P.~Zoller, A.~M. Rey, and J.~Ye.
\newblock Spectroscopic observation of su(n)-symmetric interactions in sr orbital magnetism.
\newblock {\em Science}, 345:1467, 2014.

\bibitem{zhou20223}
Xiaofan Zhou, Gang Chen, and Suo-Tang Jia.
\newblock Su (3) spin-orbit coupled fermions in an optical lattice.
\newblock {\em Chin. Phys. B}, 31(1):017102, 2022.

\bibitem{zohar2013}
Erez Zohar, J.~Ignacio Cirac, and Benni Reznik.
\newblock Cold-atom quantum simulator for su(2) yang-mills lattice gauge theory.
\newblock {\em Phys. Rev. Lett.}, 110:125304, 2013.

\end{thebibliography}

\vspace{0.5cm}
\noindent {\Large\textbf{Acknowledgements}}

DW acknowledges that this work was supported by the CQT/MoE (Grant No. R-710-002-016-271), the Singapore Ministry of Education Academic Research Fund Tier2 (Grant No. MOET2EP50220-0008), FC acknowledges support from Institut Universitaire de France, PEPR Project Dyn-1D (ANR-23-PETQ-001) and ANR International Project ANR-24-CE97-0007 QSOFT.

\vspace{0.5cm}
\noindent {\Large\textbf{Authors contributions}}

C. Madasu, K. R., and C. C. K. developed and calibrated the experimental apparatus. C. Madasu, C. Mitra, L. G. collected the data. T. Z. W., C. Miniatura, F. C., C. Madasu, and D. W. analyzed the data. All authors participated to the writing of the manuscript. D. W. supervised the work.

\vspace{0.5cm}
\noindent {\Large\textbf{Competing Interests Statement}}

The authors declare no competing interests.

\noindent {\Large\textbf{Tables}}

\begin{table}[!h]
	\centering
	\begin{tabular}{|c|c|c|c|}
		\hline
		$\ket{S}$ & $\ket{R}$ & $\ket{G}$ & $\ket{B}$\\
		\hline
		$|\braket{\psi_0|S}|^2$ & 0.06(3) & 0.94(3) & 0\\
		$\mathrm{Arg}\{\braket{\psi_0|S}\}/\pi$ & -0.53(6) & 0.0 & - \\
		\hline
	\end{tabular}
\end{table}
Table I. Initial color state prepared $\ket{\psi_0}$ after the ramp sequence protocol. The targeted initial color state is $\ket{G}$.

\noindent {\Large\textbf{Figure Legends}}

FIG. 1. \textbf{Experimental setup}. \textbf{a.} A laser beam (red lines), lying in the ($x,y$)-plane, is split into three parts and redirected on a cold atomic cloud of strontium atoms (grey disk). The polarization of each beam is indicated on the graph. Electro- and acousto-optic modulators (EOM, AOM) and quarter-wave plates (QWP) control each beam frequency, Rabi frequency, and polarization state.  A magnetic field bias $\textbf{B}=67G$ is applied along the $y$-axis. The in-plane atomic mean momentum \textbf{p} and its polar angle $\theta$ are defined in the laser beams reference frame (see Methods). 
\textbf{b.} 
The red arrows show the quasi-resonant atomic transitions of the intercombination line driven by the laser beams. These transitions form a 2-tripod scheme made of a left and a right tripod configuration sharing the common ground state $\ket{5/2}_g$. The 2-tripod laser Rabi frequencies and single photon detunings are $\Omega^{l, r}_{1,2,3}$ and $\delta^{l, r}_{1,2,3}$, respectively. Here, all the Rabi frequencies are equal, namely $\Omega^{l,r}_{1,2,3} = \Omega$.
\textbf{c.} The bright ($\ket{\textrm{B}}_j$, $j=1,2,3,4$) and color ($\ket{R}$, $\ket{B}$, $\ket{G}$) dressed states diagonalize the atom-laser coupling Hamiltonian in the rotating wave
approximation. At resonance ($\delta^{l, r}_{1,2,3} = 0$), the color states are degenerated with a zero eigenenergy and separated from the bright states by a frequency shift of the scale $\Omega$. For moving atoms and quasi-resonant beams, the color states are quasi-degenerate as long as the Doppler effect and detunings are much smaller than the Rabi frequency. 
Then, the atomic dynamics constrained to the color space are described by an SU(3) gauge field. 
\textbf{d.} Left panel: Time-of-flight (TOF) image showing the measured velocity distribution (in units of the recoil velocity $v_r$) of the atoms in the different ground states after the initial color state preparation. Right panel: From the TOF image, we extract the populations in the different ground states (blue vertical bars) and compare them to the populations expected from the targeted initial state (green vertical bars). The black error bars are the experimental standard deviation. The fidelity of our state preparation protocol is $\mathcal{F}= 0.97(2)$.

FIG. 2. \textbf{Cartan-Weyl transformation}. \textbf{(a)} Schematic of the Cartan-Weyl SU(3) ladder operators structure leading to the all-state connectivity. \textbf{(b)} Time evolution of the populations of states $\ket{R}$ (red disks), $\ket{G}$ (green disks) and $\ket{B}$ (blue disks) for the Hamiltonian $H_{GR}$ defined in \eq{CW_H}. With an initial state as $\ket{G}$, we observe a direct $\ket{G}\rightarrow\ket{R}$ population transfer. The colored solid lines are the theoretical predictions for an atom at rest and the colored-dashed curves correspond to theoretical predictions taking into account the momentum dispersion. \textbf{(c)} Same as (b) but for Hamiltonian $H_{GBR}$. We observe an indirect $\ket{G}\rightarrow\ket{R}$ population transfer mediated by $\ket{B}$. The practical implementation of the transformations is detailed in Methods.

FIG. 3. \textbf{Color-orbit dynamics}. \textbf{(a)} Temporal evolution of the atomic velocity components $v_x$ (green circles) and $v_y$ (red squares) around their mean values $\langle v_x\rangle= -0.14(2)v_r$ and $\langle v_y\rangle= 1.91(11)v_r$. The time origin corresponds to the end of the adiabatic ramp sequence transferring bare state $\ket{9/2}_g$ to a color state and the start of the evolution in the SU(3) gauge field. The plain curves are the theory prediction considering the finite momentum spread of gas. The (conserved) mean initial momentum of the gas is ${\bf p}=8 \hbar k \,\hat{{\bf x}}$. \textbf{(b)} Fourier Transformation (FT) amplitude of $v_y-\langle v_y\rangle$ obtained by Fourier transform of the time signal in (a). \textbf{(c)} Bare state populations spectral density (PSD) as indicated on each panel. 
In all panels, the plain curves are numerical predictions using \eq{CW_H} and the finite momentum dispersion of the gas, see methods. The dashed black vertical lines indicate the predicted three Bohr oscillation frequencies $4\omega_r$, $8\omega_r$ and $12\omega_r$ for an atom at rest, see methods.

\end{document}